\documentclass{pasj00}

\begin{document}

\SetRunningHead{M. Shoji et al.}{Spectroscopic Studies of Limb Spicules}
\Received{2009/05/12}
\Accepted{2010/01/05}

\title{Spectroscopic Studies of Limb Spicules. \\
I. Radial and Turbulent Velocities}

\author{Makiko \textsc{Shoji}, Takara \textsc{Nishikawa}}
\affil{Kyoto College of Economics,
       3-1 Higashinaga-cho Ooe, Nishikyo-ku, Kyoto, 610-1195}
\email{shoji@kyoto-econ.ac.jp, nishikawa@kyoto-econ.ac.jp}
\author{Reizaburo \textsc{Kitai}}
\affil{Kwasan Observatory, Graduate School of Science, Kyoto University, \\
       17 Ohmine-cho Kita Kazan, Yamashina-ku, Kyoto, 607-8471}
\and
\author{Satoru \textsc{Ueno}}
\affil{Hida Observatory, Graduate School of Science, Kyoto University, \\
       Kurabashira, Kamitakara-cho, Takayama, Gifu, 506-1314}


\maketitle

\begin{abstract}
We made high-resolution spectroscopic observations of limb-spicules in H$\alpha$  
using the Vertical Spectrograph of Domeless Solar Telescope at Hida Observatory.
While more than half of the observed spicules have Gaussian line-profiles,
some spicules have distinctly asymmetric profiles which can be fitted with two Gaussian components.
The faster of these components has radial velocities of 10--40 km~s$^{-1}$
and Doppler-widths of $\sim$0.4 \AA \ 
which suggest that it is from a single spicule oriented nearly along the line-of-sight.
Profiles of the slower components and the single-Gaussian type show very similar characteristics.
Their radial velocities are less than 10 km~s$^{-1}$ and the Doppler-widths are 0.6--0.9 \AA.
Non-thermal ``macroturbulent'' velocities of order 30 km~s$^{-1}$ are required to explain these width-values.
\end{abstract}

\section{Introduction}

Spicules are one of the fundamental elements of the quiet solar chromosphere,
so it is important to know the physical properties of spicules 
to understand the quiet Sun.
\citet{Pasachoff-Beckers} did a pioneering observation of limb spicules
with many chromospheric emission lines, 
and derived the basic morphological and spectroscopic properties.
They also suggested the rotation of spicule gas from the inclined spectra.
\citet{Krat} also made spectroscopic observations of limb spicules in 5 spectral lines,
and reported that emission-line profiles are classified into 
narrow ones and wide ones. 
However, they were not sure whether they observed single individual spicules,
since the spatial resolution of their study was several arcseconds.
Summarizing these observational works,
\citet{Beckers} reviewed the morphology and spectroscopic properties of spicules
and summarized that a spicule has a diameter of 400--1500 km 
and mean Doppler velocity of 10 km~s$^{-1}$.
\citet{Kuli-Niko} observed 650 H$\alpha$ line profiles of 25 spicules, and 
categorized them into two groups. 
One group is of relatively small intensity and narrow emission profiles,
and the other is of brighter and wider ones.
The characteristics of the latter group was thought to be due to the overlapping
of unresolved spicules.
They suggested that the H$\alpha$ spicules have temperature of 6000 K
and non-thermal velocities of $\sim$25 km~s$^{-1}$.
Follow-up spectroscopic observations of limb spicules 
under much better seeing conditions 
were done by \citet{Kuli} with the 53-cm Lyot coronagraph of 
Abastumani Astrophysical Observatory 
and by \citet{Hasan} with Sacramento Peak VTT.
From the study of line-of-sight (LOS) velocity and line-width distribution
along the axis of spicules,
they got a result that, in majority of spicules, the H$\alpha$ line-width
does not change spatially along the axes of spicules.
\citet{Kuli} suggested that the line width of the majority of spicules
are due to the overlapping of gas moving 
with relative velocities of 20--30 km~s$^{-1}$.

\citet{Nishikawa} used H$\alpha$ filtergrams to determine the diameter 
and the motion of limb spicules.
This morphological study found 
that the speed of the apparent up-and-down motion of large spicules 
is $\sim$50 km~s$^{-1}$.
Assuming ballistic motion, he estimated the initial velocity to be 
around 100 km~s$^{-1}$,
though there has been no other report of such high velocities.
The typical diameter of spicules was about 500 km, 
although the best images show thinner components of 200 and 350 km in diameter.

Recent studies using Ca\emissiontype{II} H filtergram from Hinode has shown 
that spicules move more dynamically and have much smaller diameters 
than previously thought.
\citet{DePontieu-b} found that 
the apparent maximum velocity of ``type I'' spicules is 15--35 km~s$^{-1}$, 
and the apparent upward velocity of ``Type II'' spicules is 40--300 km~s$^{-1}$.
\citet{Suematsu} reported that a spicule consists of highly dynamic multi-threads 
as thin as a few tenths of an arcseconds and shows lateral movement 
or oscillation with rotation.

Although Hinode's direct imaging provides us highly valuable images, 
we need spectroscopic observations of spicules 
with high spatial and temporal resolutions
to know the actual physical state of the spicular gas.
As \citet{Sterling} pointed out, knowledge of the actual physical properties is
essential to theoretically clarify the dynamics and ejection process of spicules
with numerical simulations. 
Recently, \citet{Pasachoff} presented the ground-based observations with 
very high spatial resolution done at the Swedish 1-m Solar Telescope on La Palma.
They analyzed the imaging observations at five wavelengths around  H$\alpha$.
Although their observation provides us a new statistics on morphological
and dynamical properties of limb spicules, 
they have derived the spectroscopic quantities such as LOS velocities 
under the rather simplified assumption of the Gaussian emission profiles.
At present, further analyses of full line profiles in detail remain to be done
to find the thermodynamical properties of spicular gas.

In our series of papers, we present the analyses of 
spectroscopic observations of limb spicules 
in H$\alpha$ with high spatial resolution, 
which is comparable to the best resolution of ground-based observations.
This paper reports and discusses the line widths and the radial velocities 
of spicules
derived from our spectral images.

\section{Observation}

The observations were made on September 8, 2005 at Hida Observatory, 
using the Vertical Spectrograph of the Domeless Solar Telescope.
The 3rd order spectra gave us the dispersion of 0.300 \AA ~mm$^{-1}$ 
and the spectral resolution 0.0123 \AA \ (which corresponds 0.56 km~s$^{-1}$) 
with the slit width 0.1 mm.
H$\alpha$ spectra were taken by Kodak Megaplus 1.6i CCD camera 
with 150 msec exposure time.
The slit position was fixed during each series of observations
and the camera continuously took the spectral images with a 1 sec time cadence. 

\begin{figure}
  \begin{center}
    \FigureFile(80mm,80mm){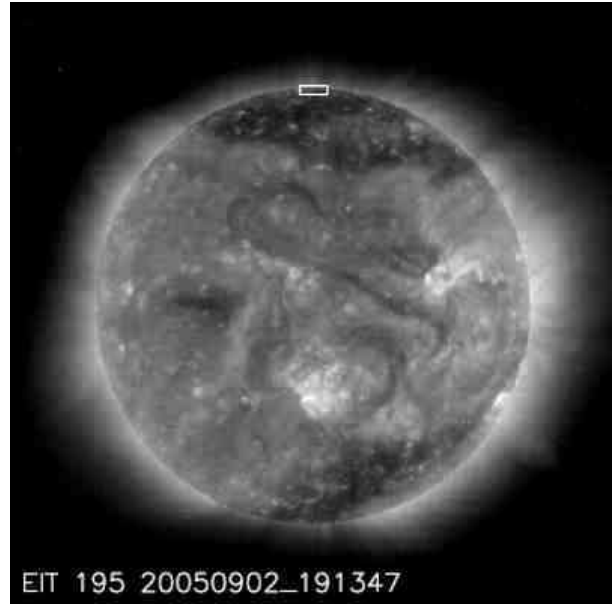}
  \end{center}
  \caption{SOHO EIT shows a coronal hole around the north pole.
  The rectangle indicates the observed region.}\label{fig:soho}
\end{figure}

\begin{figure}
  \begin{center}
    \FigureFile(80mm,65.2mm){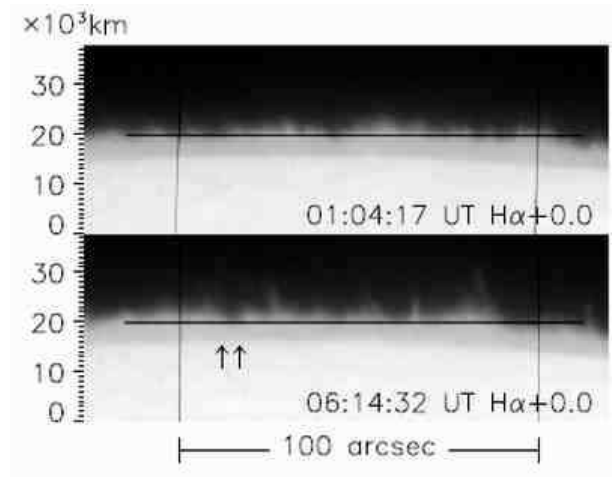}
  \end{center}
  \caption{H$\alpha$ images of the observed regions at 01 UT and 06 UT.
  The horizontal line is the spectrograph slit.
  Two vertical lines are fiducial hairlines.
  The images are presented in logarithmic intensity scale 
  to enhance the details around the slit.
  The arrows indicate the positions of two exceptional events 
  mentioned in section \ref{sec:discuss}.}\label{fig:slit}
\end{figure}

We targeted the north polar limb 
which was covered by a large coronal hole at that time 
(figure \ref{fig:soho}).
The slit was positioned parallel to the limb and at the tops of the spicules 
to get the optically thin profiles (figure \ref{fig:slit}).
We were able to obtain more than 3500 frames of H$\alpha$ spectral images 
with high spatial resolution
during two series of observations around 01 UT and 06 UT.
The height of the slit position at the nearest point from the solar limb 
seen in H$\alpha$+0.9 \AA \ 
was 4200 km for 01 UT and 3700 km for 06 UT.

\begin{figure*}
  \begin{center}
    \FigureFile(161mm,203mm){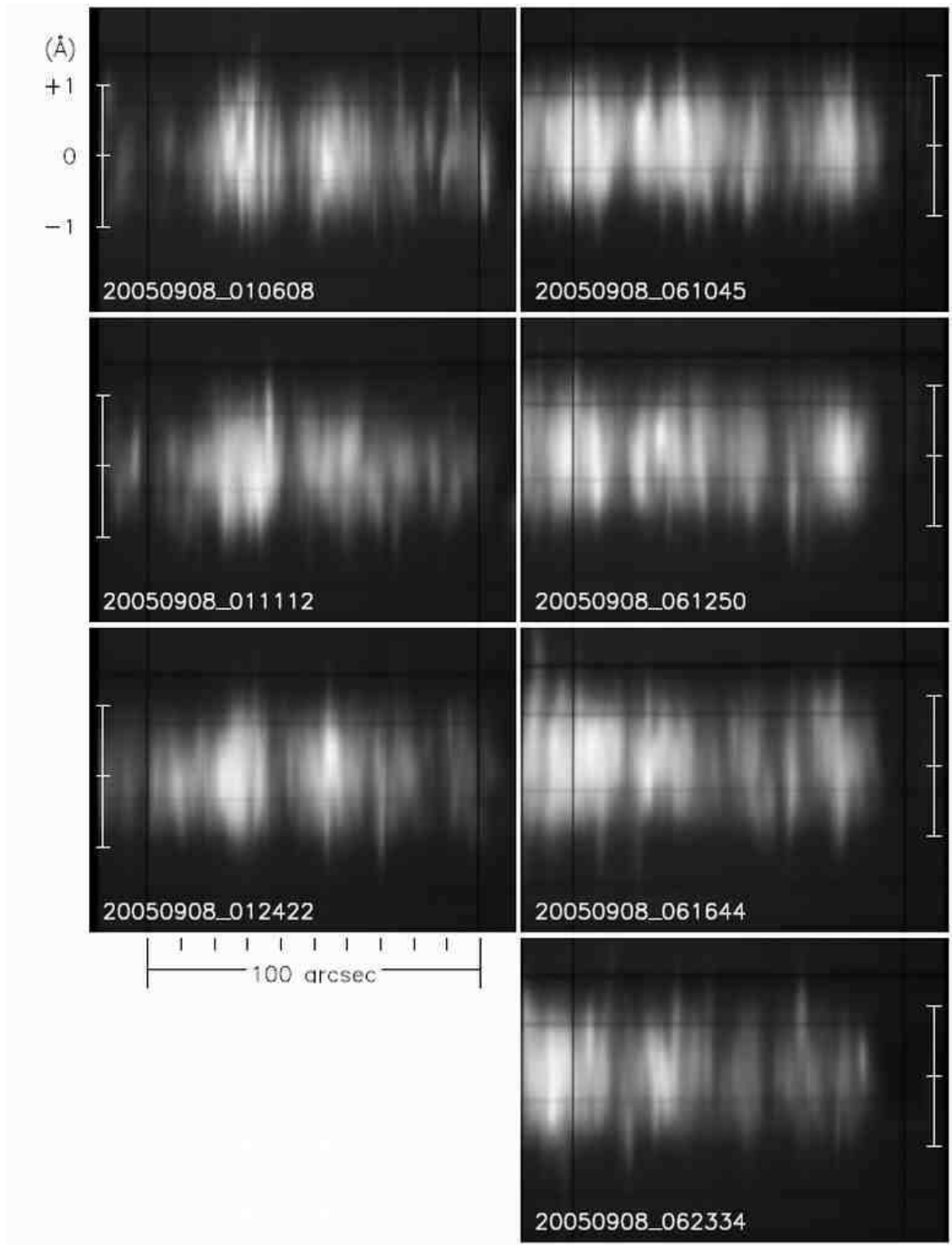}
  \end{center}
  \caption{H$\alpha$ spectra of spicules analyzed in this paper. 
  Two vertical lines correspond to the fiducial hairlines.}\label{fig:spectra}
\end{figure*}

Among the numerous spectral images, 
we first discarded the frames 
showing dark lane of self-absorption around the H$\alpha$ line center, 
since 
their profiles with flat-bottomed dip require the radiative transfer analysis to interpret
and we wish to concentrate our profile analysis to optically thin emissions.
Then, we further searched the frames showing good spatial resolution.
As a result, seven frames shown in figure \ref{fig:spectra} were
picked up for analysis,
three from the sequence around 01 UT and four from 06 UT. 
The spatial FWHM values of the sharp streaks measured at $\pm$0.8 \AA \ 
are 1000--1900 km.

Recent Hinode/SOT observations found that the spatial widths of spicules are
around 200 km or less.
The spatial resolution of our spectroscopic observation is much coarser than 
the true widths of spicules.
Unfortunately, our observation was before the launch of Hinode
and we could not make use of the high-resolution images of SOT to compare.

\section{Data Analysis}

Dark subtraction and flat fielding were applied to the observed spectral images,
then a mean sky spectrum was subtracted 
to obtain the proper emission profile of the spicules.

\subsection{Method of Profile Fitting} \label{sec:method}

We got the profiles at every pixel position along the slit 
between the hairlines in figure \ref{fig:slit}. 
The spacing is 0.15 arcsec per pixel, so we got about 650 profiles from each frame.
Telluric lines were removed from the profiles,
then each profile was fitted first with a single Gaussian curve 
as a function of wavelength from H$\alpha$ line-center ($\Delta \lambda$), 
which determines relative peak intensity $I_{0}$, Doppler width $w_{\mathrm{D}}$, 
and Doppler shift $\Delta \lambda_{\mathrm{D}} $ of the emission.
\begin{equation}
  I_{\mathrm{single}}(\Delta \lambda ) 
  = I_{0} e ^{ -\left ( \frac{\Delta \lambda - \Delta \lambda _{\mathrm{D}}}{w_{\mathrm{D}}} \right ) ^{2}}
\end{equation}

Some profiles were asymmetrical and were not well fitted with a single Gaussian curve.
The goodness of fit was eveluated by a reduced chi-square ($\chi^{2}$) statistic.
First we eliminated the low-intensity ($I_{0} < 3.0$ sky brightness) profiles 
as they had a low signal-to-noise ratio.
Then we introduced $\chi / I_{0}$ as a threshold for a good fit,
because the effect of the $\chi$-values are influenced by intensity. 
For the profiles with a large $\chi / I_{0}$ value, 
we tried to fit the observed profile with a sum of two Gaussian curves, 
and get two sets of parameters.
\begin{equation}
  I_{\mathrm{double}}(\Delta \lambda ) 
			= I_{\mathrm{slow}} e ^{- \left (\frac{\Delta \lambda - \Delta \lambda _{\mathrm{slow}}}{w_{\mathrm{slow}}}\right)^{2}}
			+I_{\mathrm{fast}} e ^{-\left(\frac{\Delta \lambda -\Delta \lambda _{\mathrm{fast}}}{w_{\mathrm{fast}}}\right)^{2}}
\end{equation}
The subscripts ``slow'' and ``fast'' denote the component 
with smaller and larger Doppler-shift values
respectively. 

\begin{figure*}
  \begin{center}
    \FigureFile(140mm,86mm){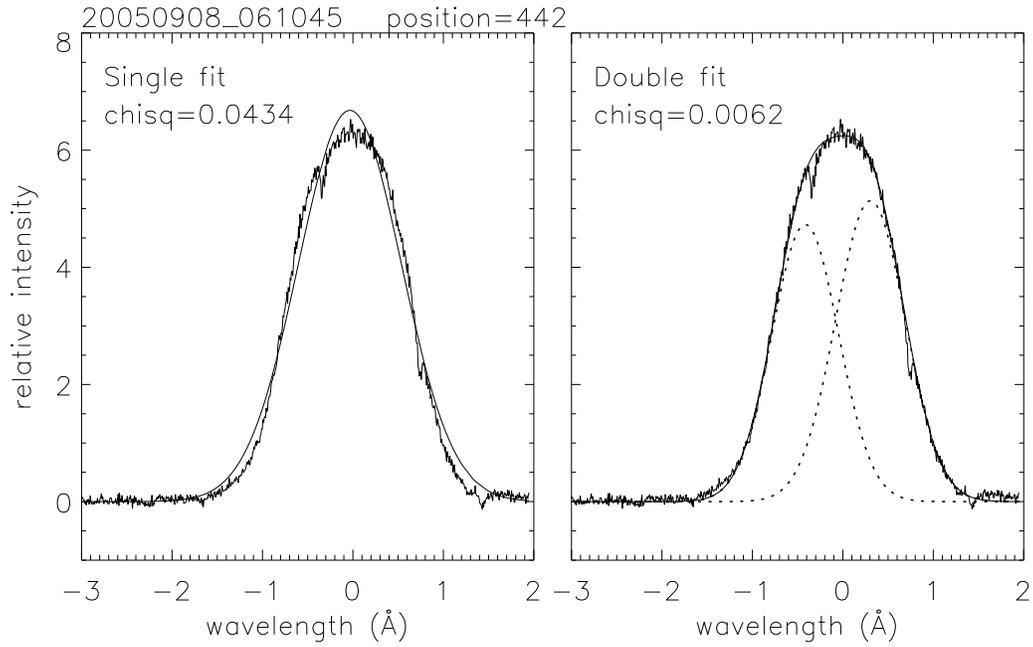}
  \end{center}
  \caption{Comparison of results with single and double Gaussian fitting.
  Though the original emission profile is almost symmetrical,
  single Gaussian fitting (left) cannot reproduce the observed profile.
  Double Gaussian fitting for the same profile (right) works well. }\label{fig:fitB}
\end{figure*}

Figure \ref{fig:fitB} is an example of 
a single-peaked and almost symmetrical profile
which requires the double Gaussian fitting. 
While there remains substantial deviations at the peak and far wing of the
profile in the case of the single Gaussian fitting (left panel), 
the double Gaussian fitting (right panel) 
can well follow the whole curve of the observed profile.
The goodness of fit (chisq) becomes 10 times better
in the case of the double Gaussian fitting. 
Similarly, for single-peaked
but fairly asymmetrical profiles, the double Gaussian fitting works very
well in our data.

All of the analyzed profiles are successfully fitted with 
single or double Gaussian curves. 
This confirms that the emissions in our selected frames were 
from optically thin part of the spicules.

\subsection{Effect of Atmospheric Image Blurring}\label{sec:deconv}

\begin{figure*}
  \begin{center}
    \FigureFile(161mm,67mm){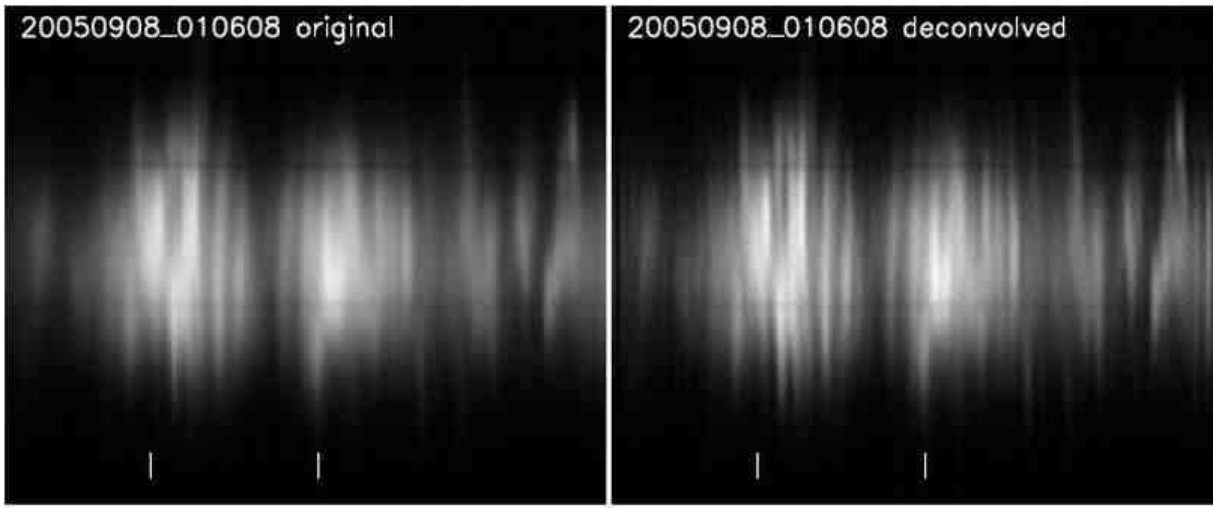}
  \end{center}
  \caption{The observed spectral image (left) and the processed image
with the Wiener deconvolution method (right).}\label{fig:filteredImg}
\end{figure*}

\begin{figure*}
  \begin{center}
    \FigureFile(164mm,170mm){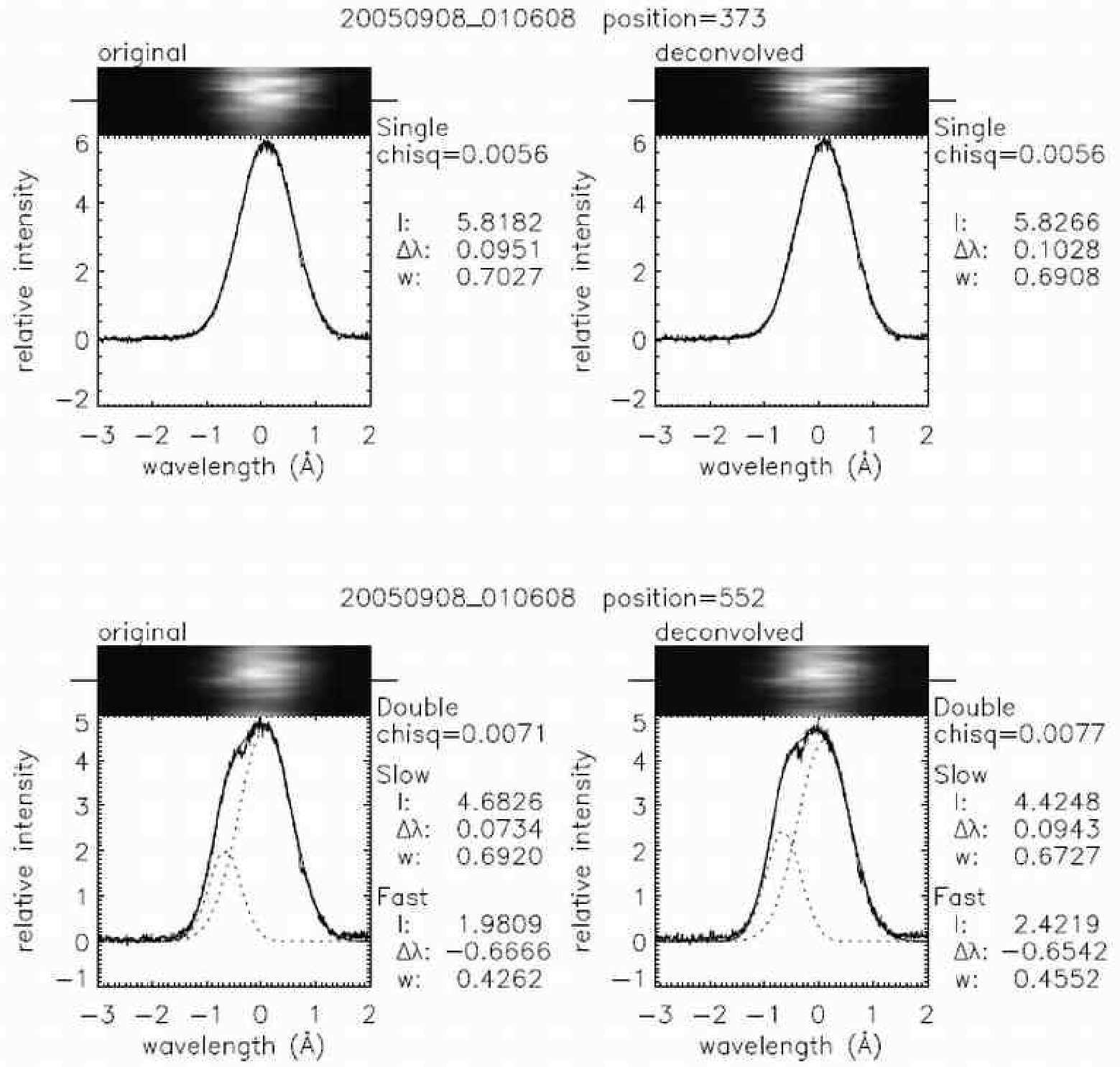}
  \end{center}
  \caption{Examples of the emission profiles to compare 
  between the original and the ``deconvolved''
  spectral images shown in figure \ref{fig:filteredImg}.}\label{fig:filteredProf}
\end{figure*}

The spatial pattern of the spicule emission is blurred by the seeing effect.
We tried to recover un-blurred spectral images with the help of 
1-D Wiener filter method and checked 
the effect of atmospheric image blurring on the spectral parameters 
of emission profiles.
We assumed that the point-spread-function (PSF) of the atmospheric blurring 
is a normal distribution. 
The width of the PSF was estimated from the smallest spatial width 
of observed emission 
and assumed to be uniform over the field of view. 
The power spectrum of the noise
was estimated from the observed intensity fluctuation 
at the continuum wavelength domain where no spicular emissions were detected.
Since its power spectrum was nearly flat,
we assumed the noise to be white-noise.

An example of compensated spectral image is shown in figure \ref{fig:filteredImg} 
with its original observed image.
We can see that the emission streaks became sharper in spatial direction
and more clear in the ``deconvolved'' image.

The same spectral fitting method as described in the subsection \ref{sec:method} 
was applied to the deconvolved image
to see the effect of atmospheric blurring on spectroscopic quantities.
Figure \ref{fig:filteredProf} shows the results for two typical profiles.
We can see that the spectral parameters do not change so much 
between the originally observed and the deconvolved image, 
except the intensity of the fast component. 
We performed the same comparison on all the profiles in the deconvolved spectral image, 
and estimated the effect of the atmospheric blurring on the spectroscopic parameters.
For the single-Gaussian fitted profiles,
$I_{0}$ changes less than $\pm$5\%, and $w_{\mathrm{D}}$ changes less than $\pm$1\%.
The change of $\Delta \lambda_{\mathrm{D}} $ is smaller then 0.05 \AA.
In the case of asymmetric profiles, 
the changes of $w_{\mathrm{slow}}$ and $\Delta \lambda_{\mathrm{slow}}$ 
are also very small.
$w_{\mathrm{fast}}$ changes $\pm$8\%, and 
$\Delta \lambda_{\mathrm{fast}}$ changes $\pm$6\%.
$I_{\mathrm{slow}}$ decreases 4.5\%, with 9.6\% deviation.
Only $I_{\mathrm{fast}}$ shows significant change, with average increase of about 14\%.
Increase in $I_{\mathrm{fast}}$ is expected since the deconvolution 
converges the blurring of isolated emission.  
Thus the spectroscopic parameters, 
especially the widths and the shifts of fitted components 
are not so influenced by the atmospheric blurring of the images.

Our method of seeing correction is based on a few
assumptions, which are 
(1) the uniformity of seeing  over the field of view,
(2) whiteness of the power of background noise,
etc.
In the deconvolved image, there appear faint and equally-spaced
emission streaks in the featureless part of the original spectral image.
These equally-spaced emission patterns are probably artificial due to the 
drawback of the method we used.
So we will proceed our discussion based on the analysis 
without the correction of atmospheric blurring.

\section{Results}

\begin{figure*}
  \begin{center}
    \FigureFile(142mm,169mm){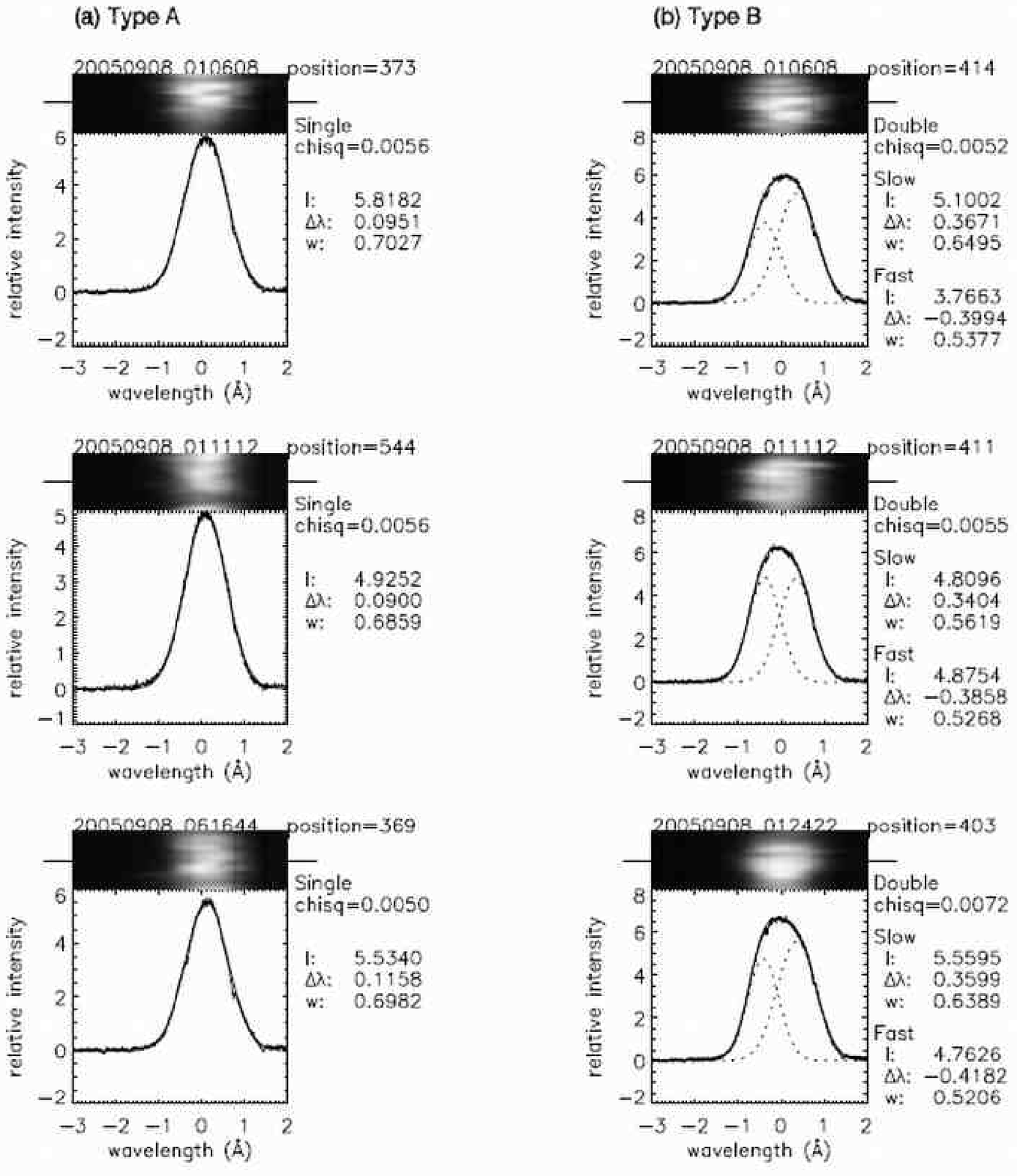}
  \end{center}
  \caption{Examples of the line profiles. 
  The unit for wavelength shift ($\Delta \lambda $) and Doppler width ($w$) values 
  is \AA. 
  (a) Typical Type A profiles, which are well fitted with a single Gaussian curve. 
  (b) Typical Type B profiles, which are symmetrical, but cannot be fitted well 
  with a single Gaussian curve.}\label{fig:typeAB}
\end{figure*}

\begin{figure*}
  \begin{center}
    \FigureFile(132mm,212mm){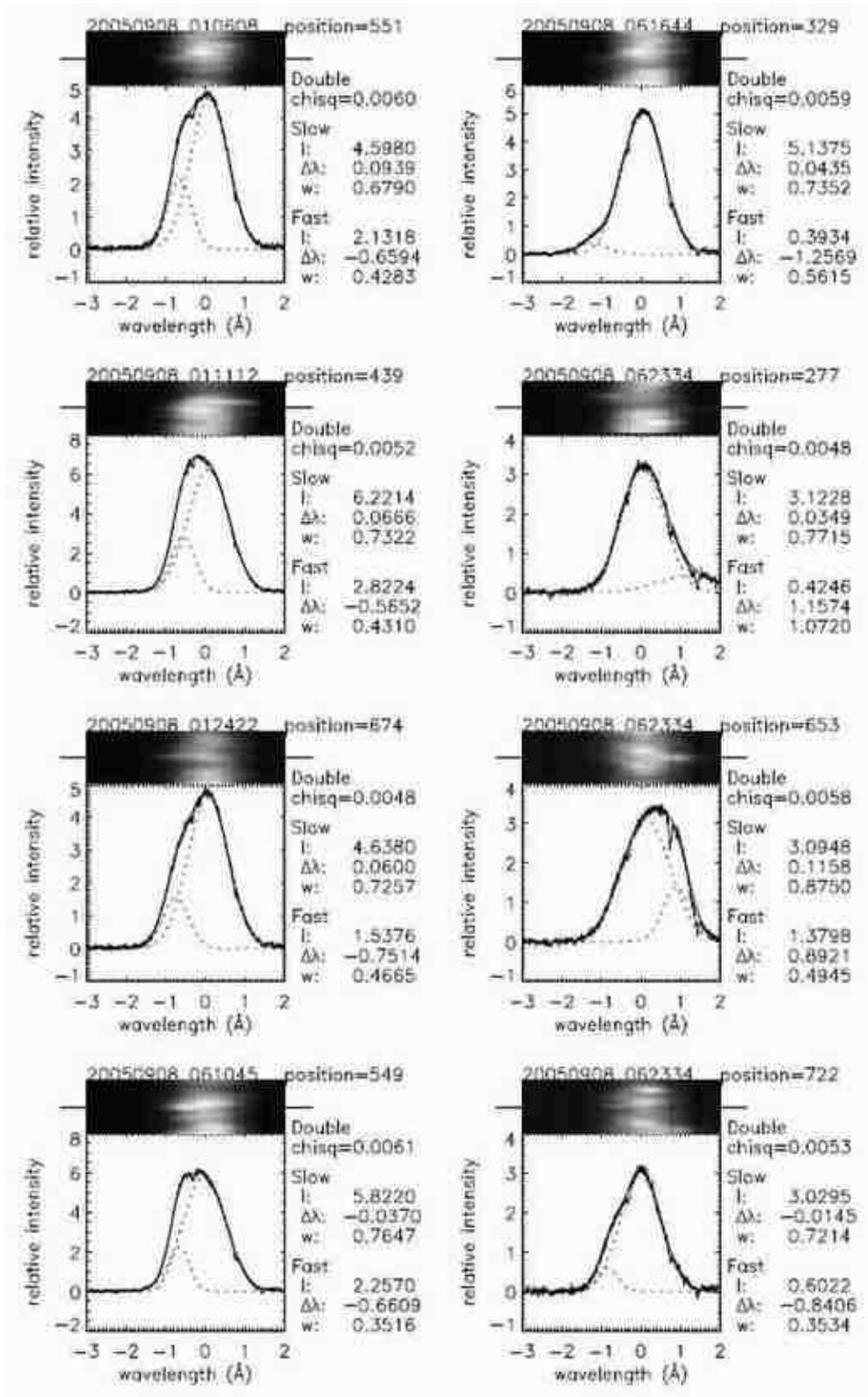}
  \end{center}
  \caption{Examples of the typical line profiles of Type C, 
  which are asymmetrical and separated to slow and fast Gaussian components. 
  The unit is the same as figure \ref{fig:typeAB}.}\label{fig:typeC}
\end{figure*}

\begin{table}
  \caption{Classification of H$\alpha$ profiles.
  Number of points where the line profile is categorized into three types 
  are tabulated.
  Type A: well fitted with single Gaussian profile.
  Type B: ill fitted with single Gaussian, but almost symmetrical. 
  Type C: asymmetrical profile which can be separated into slow and fast components.
  }\label{tbl:type}
\begin{center}
\begin{tabular}{cccc}
  \hline
  Time (UT) & Type A & Type B & Type C \\     
  \hline
  01:06:08 & 226 & ~96 & ~48 \\
  01:11:12 & 234 & 149 & ~75 \\
  01:24:22 & 397 & ~78 & ~36 \\
  06:10:45 & 369 & 161 & ~53 \\
  06:12:50 & 351 & 111 & ~69 \\
  06:16:44 & 253 & 103 & 137 \\
  06:23:34 & 233 & 128 & ~63 \\
  Total & 2063 (61\%) & 826 (25\%) & 481 (14\%) \\
  \hline
\end{tabular}
\end{center}
\end{table}

\begin{figure*}
  \begin{center}
    \FigureFile(131mm,182mm){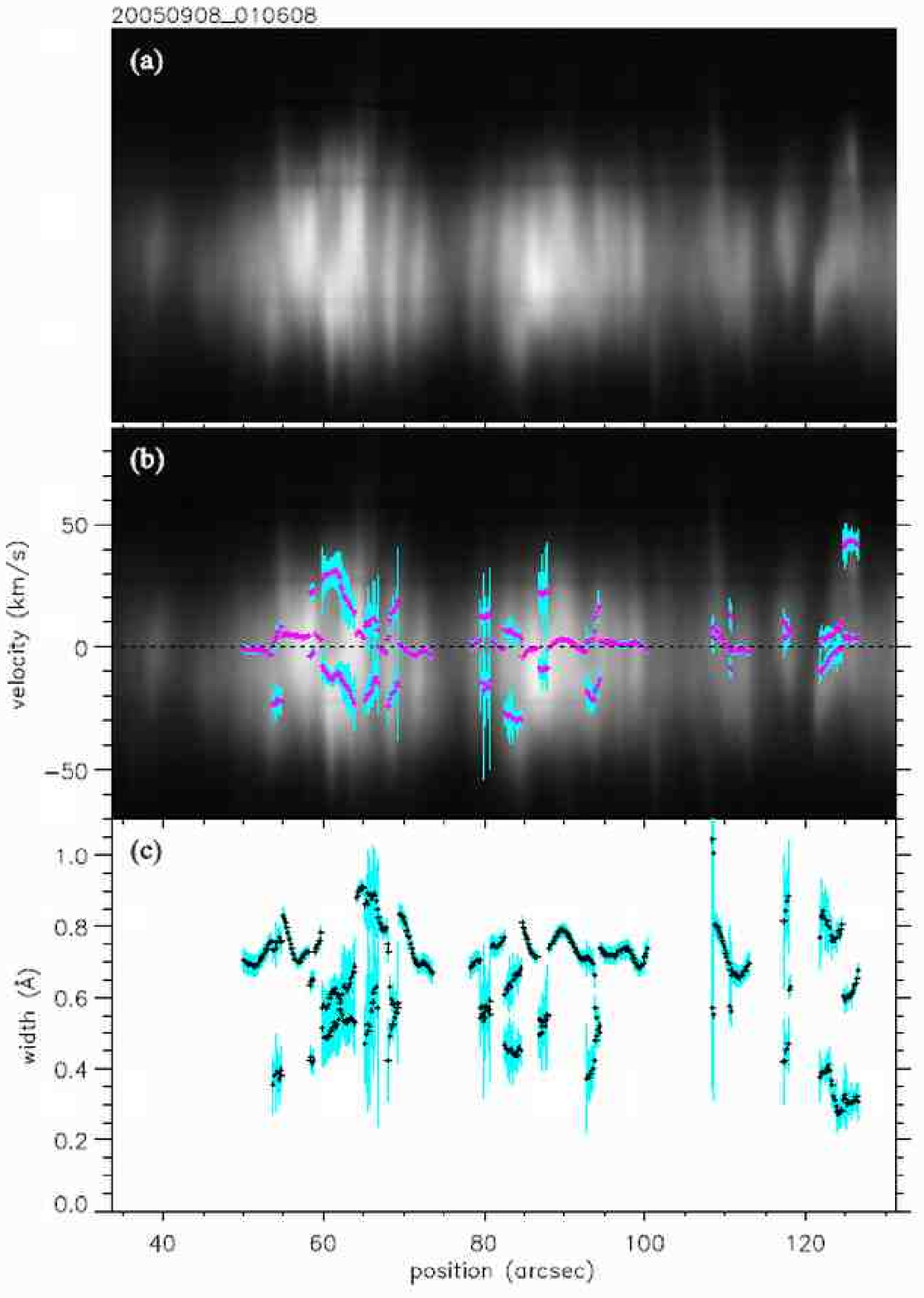}
  \end{center}
  \caption{(a) An example image showing Doppler shifted streaks of spicules.
  (b) Radial velocities determined by single or double Gaussian fitting are plotted 
  over the same image as (a). Error bar (light blue) indicates the standard deviation 
  of each velocity value. 
  (c) Doppler widths of the profiles at the same positions as (b).
  Note that two values are plotted at the same axis position 
  when the profile is fitted with double Gaussian curves.}\label{fig:example}
\end{figure*}

Line profiles of spicules can be grouped into three types.
Profiles of the first type (Type A) can be fitted well with a single Gaussian curve.
The second type (Type B) is not well fitted with single Gaussian, 
but it is nearly symmetrical around the H$\alpha$ line center (figure \ref{fig:fitB}).
The last type (Type C) is asymmetrical, 
and separated into an almost stationary component 
and a highly shifted component.
Note that the difference between Types B and C is actually a matter of degree,
and there are many profiles in-between.
For argument's sake, we take the profiles with 
$|\Delta \lambda _{\mathrm{fast}}/\Delta \lambda _{\mathrm{slow}}|>4.0$ 
as Type C and treat all other double Gaussian profiles as Type B.
Figures \ref{fig:typeAB} and \ref{fig:typeC} show the examples of 
typical profiles of each type,
and table \ref{tbl:type} lists the number of points categorized into each type.

\begin{figure}
  \begin{center}
    \FigureFile(60mm,213mm){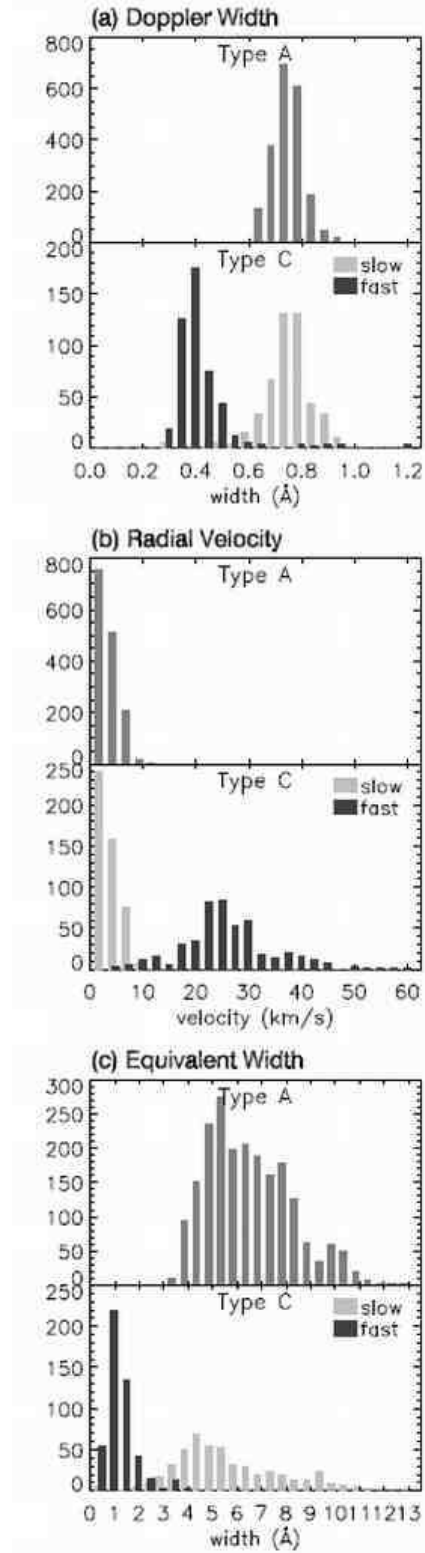}
  \end{center}
  \caption{Histograms made from the results of all frames. 
  ``Slow'' and ``fast'' mean the slow and the fast components of the Type C profile.  
  (a) Doppler width.  (b) Radial velocity (absolute value). 
  (c) Equivalent width.}\label{fig:hist}
\end{figure}

Figure \ref{fig:example} is an example showing estimated LOS velocities and widths 
with the original spectral image.  
In figure \ref{fig:hist}, we summerize the distribution of Doppler widths, 
radial velocities, 
and equivalent widths in histogram format.

\subsection{Doppler Width} \label{sec:width}

Figure \ref{fig:hist}a shows that the Doppler widths 
of the Type A profile ($w_{\mathrm{D}}$) 
are 0.6--0.9 \AA.
Widths of the slow components of the Type C profile ($w_{\mathrm{slow}}$) 
are also 0.6--0.9 \AA, 
while widths of the fast component ($w_{\mathrm{fast}}$) are 0.3--0.6 \AA.

To identify the line-broadening mechanism,
we calculated the theoretical Doppler width of H$\alpha $ profile 
with given temperature $T$, microturbulence $\xi $, and macroturbulence $\Xi $,
\begin{equation}
  w=\frac{\lambda }{c}\sqrt{\frac{2kT}{m}+\xi ^2+\Xi^2}
\end{equation}
where $\lambda $ is the wavelength of H$\alpha$ line, $c$ is the light velocity,
$k$ is Boltzmann constant, and $m$ is hydrogen mass.

\begin{figure}
  \begin{center}
    \FigureFile(80mm,95mm){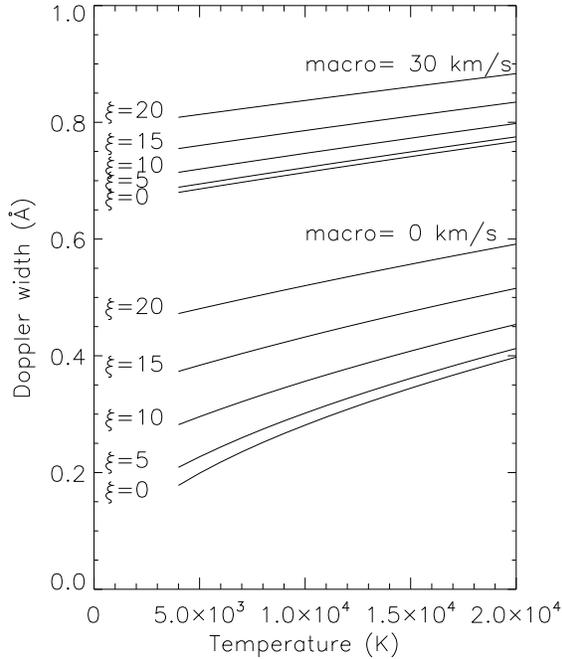}
  \end{center}
  \caption{Theoretical Doppler width, 
  calculated as a function of temperature with different values of 
  microturbulent velocity $\xi $ (km s$^{-1}$)
  and macroturbulent velocity.}\label{fig:calc}
\end{figure}

The result is shown in figure \ref{fig:calc}.
When $\Xi =0$, the observed value $w_{\mathrm{fast}}=0.35$ \AA \ is found 
within a reasonable range 
of temperature and microturbulence, $T\simeq 15000$ K and $\xi\simeq 5$ km~s$^{-1}$.
On the other hand, even if we assume as high a temperature as $T=20000$ K
with microturbulence $\xi=20$ km~s$^{-1}$ 
we cannot get a width more than 0.6 \AA.
While \citet{Beckers} listed the most likely temperature as 9000--16000 K,
\citet{Makita} derived ionization temperature for hydrogen to be 5200 K 
and excitation temperature for H$\alpha$ to be 5020 K, 
and other authors also reported lower temperatures.
Therefore, it is unlikely that the tempetrature of spicules to be higher than 20000 K. 
And considering that microturbulent velocity cannot be supersonic,
we need additional ``macroturbulence'' to explain the observed
 $w_{\mathrm{slow}}$ and $w_{\mathrm{D}}$ .

From figure \ref{fig:calc}, we find 
that the required $\Xi $ for $w_{\mathrm{slow}}$ and $w_{\mathrm{D}}$ 
is about 30 km~s$^{-1}$.
Here we use the term ``macroturbulence'' to include any non-thermal random velocities 
other than microscopic turbulence.
There exists many spicules with various LOS velocities 
along the LOS of observation.
Superposition of their emissions will result in the broadening of the observed profile.
``Macroturbulent velocity'' is a measure of the dispersion of their LOS 
velocities.

\subsection{Radial Velocity}

Figure \ref{fig:hist}b is the histogram of radial (line-of sight) velocities.
With our definition for Type C, 
the distribution of the fast component's parameters hardly overlaps 
with that of the slow component's.
Radial velocities derived from Type A profiles are less than 10 km~s$^{-1}$,
and the profiles of the slow component of Type C give similar values.
The distribution of the fast component peaks at around 25 km~s$^{-1}$
and there is also a small peak at around 40 km~s$^{-1}$.

\subsection{Correlation between Width and Velocity}

\begin{figure}
  \begin{center}
    \FigureFile(80mm,125mm){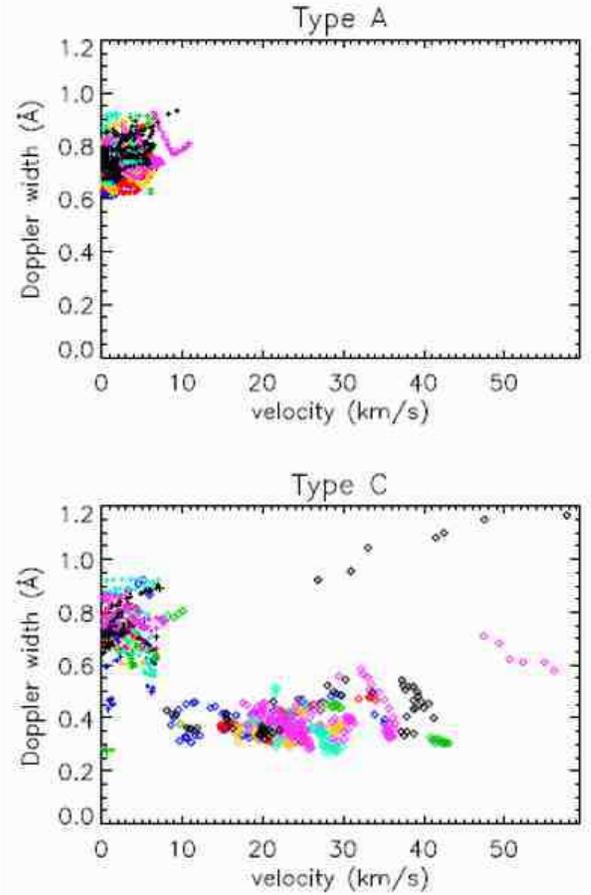}
  \end{center}
  \caption{Correlation between velocity (absolute value) and Doppler width.
  Crosses are for the slower component 
  and diamonds are for the faster component.
  Different color represents different time.}\label{fig:corr}
\end{figure}

Figure \ref{fig:corr} is a correlation plot 
between radial velocities and Doppler widths 
for Types A and C.
Again, it is clear that the slow component of the Type C profile 
has a similar distribution 
to the Type A profile.
The velocity of the fast component spreads from 10 km~s$^{-1}$ to 40 km~s$^{-1}$, 
but its width is independent of radial velocity.

\section{Summary and Discussion}\label{sec:discuss} 

We made high-resolution spectroscopic observations of limb spicules in H$\alpha$.
While more than half of the observed spicules have Gaussian line profiles (Type A),
some spicules have distinctly asymmetric profiles which can be fitted with
two Gaussian components (Type C).
The faster component of Type C profiles have radial velocities of 10--40 km~s$^{-1}$
and Doppler width of $\sim$0.4 \AA .
This width can be explained 
with temperature $\simeq$15000 K and microturbulence $\simeq$5 km~s$^{-1}$.
On the other hand, widths of Type A profiles 
and the slower component of Type C profiles
are 0.6--0.9 \AA.
Macroturbulent velocities of order 30 km~s$^{-1}$ are required 
to explain this large width.

Now we will discuss the physical interpretations of these results.

Figure \ref{fig:slit} shows that the slit-line intersects many spicules,
and it is natural that the LOS intersects several spicules, too.
Though it is difficult to observe a single spicule 
without superposition of other spicules,
Type C profile enables us to separate the emission of single spicule 
as the fast component of double-Gaussian fitted components.
Line width of the fast component can be explained 
with temperature $\simeq$15000 K and microturbulence $\simeq$5 km~s$^{-1}$.
Broad width of the Type A profiles suggest that 
many spicules along the LOS contribute to the observed emission.
Figure~\ref{fig:hist}c shows that their equivalent widths are 
five times or more larger than those of the fast component.

The LOS velocity 20--40 km~s$^{-1}$ of the fast component is consistent 
with the velocity of the apparent up-and-down motion of spicules 
reported by many authors
(e.g. \cite{Nishikawa}; \cite{DePontieu-b}; \cite{Pasachoff}). 
Because most spicules are nearly vertical to the solar limb \citep{Beckers},
their LOS velocities are much smaller than their actual velocities.
We suppose that the profiles of Type A and slow component of Type C 
are from these nearly vertical spicules,
and that the fast component is from the few spicules 
which are almost parallel to the LOS.

As shown in section \ref{sec:width}, we need additional 
macroturbulent i.e.\ non-thermal random velocities of order 30 km~s$^{-1}$
to explain the Doppler width of the slow component of Type C  
and the Type A profiles.
Other observations also reported the high non-thermal broadening.
For example, 
\citet{Makita} suggested that spicules have a turbulence of $\sim$20 km s$^{-1}$ 
in the active region,
based on the analysis of Ca\emissiontype{II} H and K profiles in the flash spectrum of
the 1958 eclipse.
And \citet{Mariska} reported average non-thermal velocity of 28 km~s$^{-1}$ 
in EUV line profiles
at the quiet limb.
Next we proceed to discuss the source of macroturbulence.

One of the possible sources is ejecting motion
of spicules inclined toward or away from the observer.
Variations in speed and inclination of ejection 
produce the dispersion of LOS velocities.
As discussed above, most spicules have small inclinations 
and so their LOS velocities 
will be much smaller than 10--40 km s$^{-1}$.
Distributions of apparent inclinations reported by \citet{Pasachoff} 
have peaks at 10 and 25 degrees, 
in which case the LOS velocity of ejection speed 40 km s$^{-1}$
will be at most 7 and 17 km s$^{-1}$.
They are too small to explain the observed widths.
There is also the height difference of ejection speed.
The spicules nearer to us or farther than the limb of the sun 
would be observed in superposition against lower heights of spicules 
whose bases are on the limb. 
However, the speed at the tops is no faster than the lower part
 (\cite{Sterling}; \cite{DePontieu-b}), 
and it will contribute little to the line broadening, 
similar to the argument above.  

Another possible source is the lateral motion of spicules
due to the Alfv\'en wave disturbance 
recently found in Hinode/SOT Ca\emissiontype{II} H filtergrams
(\cite{DePontieu-a}; \cite{Suematsu}; \cite{He}),
or the kink wave observed by \citet{He2}.
\citet{DePontieu-a} showed that the distribution 
of transverse displacements of the spicules
agrees with the velocity amplitudes around 20 km~s$^{-1}$,
while \citet{He} reported the velocity amplitude of high-frequency Alfv\'en waves
to be 4.7--20.8 km~s$^{-1}$.
Velocity amplitude of the kink wave reported by \citet{He2}
is less than 8 km~s$^{-1}$.
Superposition effects of these motions can broaden the line profiles.
Because spicules are believed to be ejected along magnetic field lines,
transverse Alfv\'en waves
broaden the H$\alpha$ line profiles of Type A or slow component of Type C.
MHD simulation also showed that the amplitude of Alfv\'en waves 
at the chromospheric height
to be  $\sim$20 km s$^{-1}$ (\cite{Suzuki}).
However,  
when the sinusoidal motions of independently disturbed spicules 
are superposed, 
its standard deviation will be $1/\sqrt{2} $ of amplitude.
Thus oscillation of velocity amplitude 20 km s$^{-1}$ will contribute
only 14 km~s$^{-1}$ to the macroturbulent velocity.
As for the fast components, we assume 
that their surrounding magnetic field is nearly parallel to the LOS, 
so the Alfv\'en waves will not contribute to line-broadening, 
agreeing with their narrow widths even if 
they are composed of multiple finer threads.
\citet{Kuli} reported the oscillations of the radial velocities 
of limb spicules
with the period of 3--7 min.
However, their amplitudes were less than 10 km s$^{-1}$,
which are too small to contribute to our observed line-widths.

Even when we include superposition effects of ejecting motions 
and Alfv\'en wave broadening,
it is still insufficient to explain the macroturbulent velocities 
of $\sim$30 km~s$^{-1}$.
So there must be some additional but unidentified 
sources for the non-thermal broadening of spicule emission profiles.
Alfv\'en waves with higher frequencies and shorter wavelengths
than those observed until now
may be present and contribute to the broadening.

Let us mention two interesting events found in our analysis.
Their characteristics are different from the common spicules discussed so far.
While most of radial velocities are less than 40 km~s$^{-1}$,
these two cases have much larger ( $>$ 45 km~s$^{-1}$) radial velocities 
as seen in the upper right area of figure~\ref{fig:corr}.
Their Doppler widths are exceptionally large, too.
They are in the spectra at 06:16:44 UT and 06:23:34 UT,
and their faint and broad profiles can be found in figure~\ref{fig:typeC}.
The positions of these events are marked in figure \ref{fig:slit}
where there is a gap in the bush of spicules,
and where the slit is the nearest to the limb.
We need different broadening mechanism for these events.
It may be an event with a magnetic reconnection, 
or something related to the spicule formation.
This phenomenon may correspond to the Type II spicules of \citet{DePontieu-b}
or to macrospicules (\cite{Bohlin}; \cite{Yamauchi}) 
or to polar-jets (\cite{Shibata}; \cite{Shimojo}; \cite{Savcheva}).
The relations are yet to be ascertained in future studies.

In this paper, we treated the results statistically without regard 
to their positions.
However, figure \ref{fig:spectra} shows that some spicules appear
as tilted streaks.
This is possibly an evidence of rotation of spicules suggested 
by \citet{Pasachoff-Beckers},
and its spatially blurred emission may contribulte to broaden the 
line profiles.

Some physical values are not determined by the H$\alpha$ profile alone.
In order to study the true origin of macroturbulence or of the high-velocity events,
we need simultaneous spectroscopic observations of spicules in multiple spectral lines,
which can be achieved with the Horizontal Spectrograph at Hida Observatory.
We also need high spatial-resolution images of Hinode/SOT
simultaneously obtained with the spectroscopic observation
to understand the details of spicules.
We will plan our next observation to accommodate to those requirements.

\bigskip

We thank the referees for the greatly helpful comments.
We are grateful for the use of EIT data obtained on the SOHO spacecraft. 
SOHO is a project of international cooperation between ESA and NASA.
The authors are supported by a grant-in-aid for the Global COE program
``The Next Generation of Physics, Spun from Universality and Emergence''
from the Ministry of Education, Culture, Sports, Science and Technology (MEXT)
of Japan, and by the grant-in-aid for ``Creative Scientific Research
The Basic Study of Space Weather Prediction'' (17GS0208, PI: K. Shibata)
from the Ministry of Education, Culture, Sports, Science and Technology of Japan,
and also partly supported by the grant-in-aid from the Ministry of
Education, Culture, Sports, Science and Technology of Japan (No.19540474).



\end{document}